# Days Alive at Home vs Out of Hospital: Why the Difference Matters for Trial Endpoints

**Authors:**

Letao Yuan[1*], Sarah N. Dawson[1,2*], David S. Robertson[1], Yue Lan[3], Andrew A. Klein[4‡], Dominique-Laurent Couturier[1,2†], Mia S. Tackney[1†], Sofía S. Villar[1,2†]

* These authors contributed equally as joint first authors.

† These authors contributed equally as joint senior authors.

‡ This author contributed in their clinical role as the NOTACS principal investigator.

**Affiliations:**

1 MRC Biostatistics Unit, University of Cambridge, Cambridge, United Kingdom

2 Papworth Trials Unit Collaboration, Royal Papworth Hospital, Cambridge, United Kingdom

3 University of Cambridge, Cambridge, United Kingdom

4 Department of Anaesthesia, Royal Papworth Hospital, Cambridge, United Kingdom

**Correspondence:** Letao Yuan, MRC Biostatistics Unit, University of Cambridge, Cambridge, United Kingdom CB2 0SR, letao.yuan@mrc-bsu.cam.ac.uk

**Abstract:**

Days Alive and Out of Hospital (DAOH) and Days Alive at Home (DAH) are increasingly used as patient-centred primary outcomes in perioperative trials, and a recent editorial has advocated DAOH for its simplicity and reliance on routinely collected data. We argue that the choice between these endpoints affects power, treatment effect estimation and missing data, and should be made at the design stage with these consequences in mind. By treating days in nursing homes or rehabilitation facilities as equivalent to days at home, DAOH assumes that being out of hospital is a valid surrogate for good recovery, an assumption that patient perspectives, including those from the NOTACS trial, call into question. Choosing DAH over DAOH therefore involves a trade-off between measuring what matters to patients more accurately and the added burden of tracking discharge destination, including a greater risk of missing data. If an intervention affects only length of stay, readmission or mortality, the two endpoints yield the same expected treatment effect, and the extra burden of tracking discharge destination brings no benefit. If it instead enables more patients to return directly home rather than to a care facility, DAOH may conceal this benefit and lose statistical power. We propose a baseline-adjusted DAH, which counts only days spent in a setting representing an escalation of care relative to the patient's baseline, as a better proxy for recovery that also accommodates hospital-at-home and virtual ward services.

In a recent editorial, Castro and colleagues review how Days Alive and Out of Hospital (DAOH) has become established as a primary outcome in perioperative research[1]. They argue that DAOH reflects outcomes that matter to patients, namely survival and departure from hospital, while being easily extracted from routinely collected data. They also note that DAOH simplifies data collection compared to Days Alive at Home (DAH) by removing the need to differentiate home from other discharge locations.

We welcome the authors' focus on these patient-centred endpoints and value their critical perspective. However, their characterisation of DAH as a precursor to DAOH overlooks the parallel evolution of these metrics outside of surgical research, where DAOH has been utilised since the early 2000s[2,3]. When Myles et al. pioneered DAH in perioperative research in 2017[4], they purposefully chose "days at home" to capture a crucial nuance: discharge to an intermediate care facility may undermine the prompt return home that patients prioritise. Consequently, neither measure should be viewed as an evolutionary successor to the other; rather, the choice between them involves a methodological trade-off that must be tailored to the specific clinical trial, as well as a deeper tension between precision and accuracy of measuring what matters most to patients versus the real-world logistics and data collection burden.

We agree that DAOH simplifies data collection by counting days out of the hospital regardless of the discharge destination. However, this approach assumes that being "out of hospital" is an acceptable surrogate for a good recovery—an assumption that is questionable in many perioperative trials. A day spent in a nursing home or rehabilitation facility reflects a vastly different state of recovery than a day spent at home, and treating them identically can lead to misleading trial results. In Figure 1, we illustrate how patients A-F can have an identical DAOH value of 25 days, while their DAH values range from 0 to 25, depending on the time spent in a nursing home (the baseline-adjusted DAH shown in Figure 1 is discussed later). This distinction matters not only clinically but also from the patient's perspective. In the NOTACS trial, patient and carer feedback following the pilot study informed the choice of DAH over length of hospital stay or DAOH: patients regarded an escalation in support, such as admission to nursing care, as an unfavourable outcome[5,6]. Similarly, in the qualitative study Castro et al.[1] cite in support of DAOH, patients explicitly identified home as the environment signifying normalcy, comfort and healing rather than simply being outside an acute hospital ward[7]. Moreover, in some settings, patients are discharged to long-term facilities to release acute beds rather than because they are ready to return home[8]. By counting institutional and home days alike, DAOH has poorer resolution than DAH and introduces potential systematic bias.

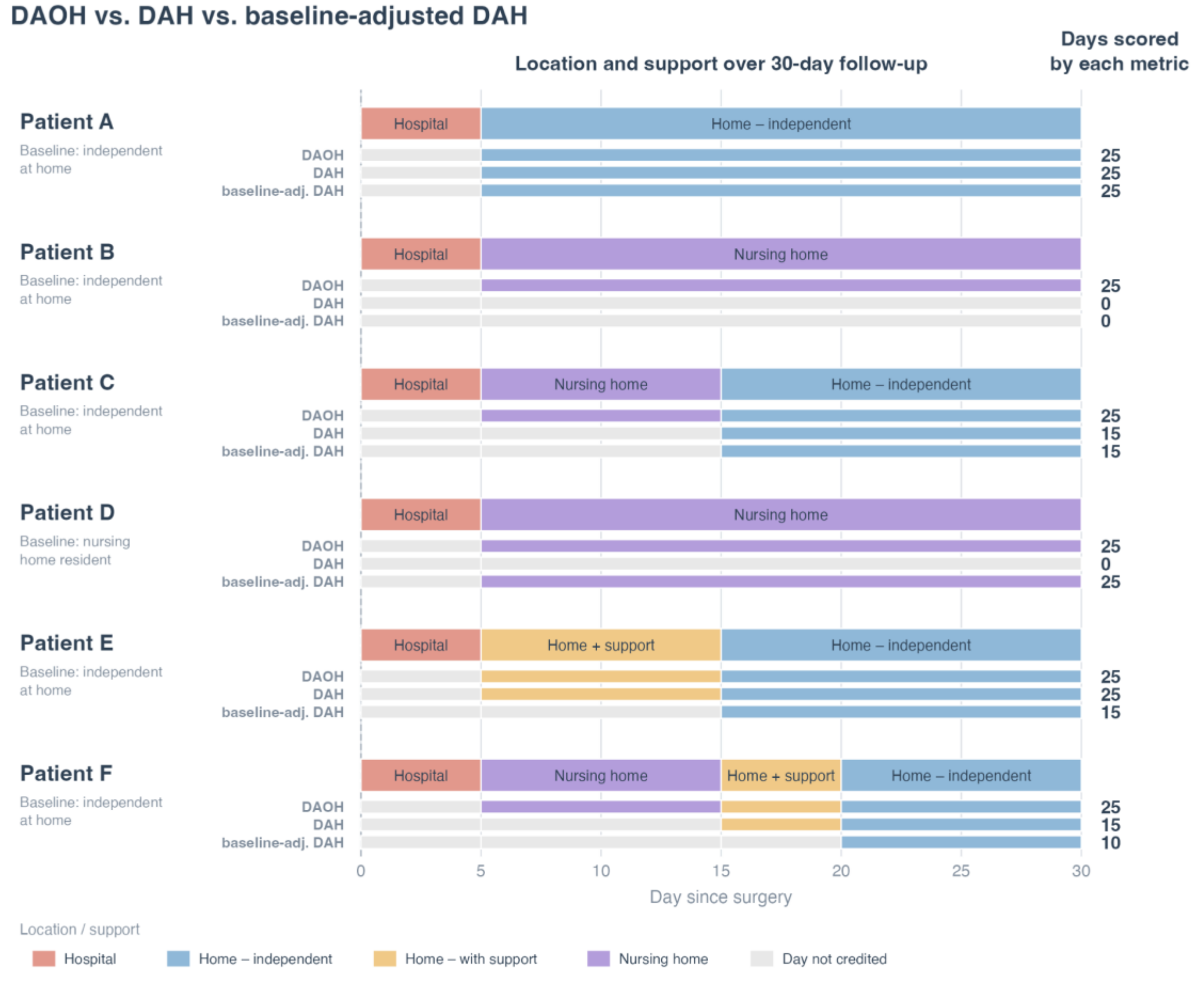

**Figure 1:** Comparing DAOH, DAH and baseline-adjusted DAH (as used in the NOTACS trial[5]) in six hypothetical patients

The choice between these two endpoints is far from trivial. First, it requires careful consideration of whether simply being out of the hospital serves as a valid surrogate for achieving a genuinely good clinical outcome. Second, if this surrogacy does not hold, investigators face a challenging trade-off: either commit to the substantially higher data collection effort required to track true post-discharge destinations—which introduces the risk of missing location data—or rely on DAOH as a proxy, accepting the potential loss of precision and introduction of bias. Furthermore, post-discharge data are inherently less reliable than hospital administrative records and are more likely to be missing via a process that is plausibly outcome-dependent, as patients with poorer recoveries are often harder to follow up.

The choice between these endpoints carries direct and potentially dramatic consequences for statistical power and treatment effect estimation. Because DAH and DAOH differ primarily in how days spent in intermediate care facilities are scored, their relative performance depends on *where* along the recovery pathway a treatment effect manifests. Consider two scenarios. First, if an intervention acts solely on hospital length of stay, readmission, or mortality, with no between-arm difference in discharge destination, the two endpoints yield the same expected treatment effect. In this case, tracking post-discharge locations adds no statistical value while increasing the data collection burden and the likelihood of missing data. Second, if the intervention shifts the discharge pathway itself—for example, by enabling more patients to return directly home rather than entering a non-acute care setting such as a nursing home or rehabilitation facility—this treatment effect could go entirely unnoticed under DAOH, as the metric dilutes or conceals this benefit, reducing statistical power.

Finally, we propose an alternative that moves beyond the binary choice between DAOH and standard DAH: a baseline-adjusted DAH, in which DAH is scored relative to each patient's baseline residence and support needs, meaning days are only subtracted if spent in a setting representing an escalation in care relative to baseline (Figures 1 and 2), as successfully implemented in the NOTACS trial[5]. Where an intervention is expected to influence post-discharge care, this approach captures the treatment effect without compromising data sensitivity—making DAH a much stronger proxy for true recovery, albeit at a non-negligible data collection cost. For example, for patient E in Figure 1, while DAH and DAOH have the same value, time spent at home with escalated support compared to baseline is uniquely captured by the baseline-adjusted DAH. This is increasingly relevant given the growth of hospital-at-home and virtual ward services, which allow patients to be discharged early while continuing to receive monitoring and treatment at home. Under DAOH and standard DAH, such days are indistinguishable from unsupported recovery at home. The baseline-adjusted DAH instead counts them as an escalation in care and thereby preserves the link between the metric and true recovery. Arguably, in settings where hospital-at-home and virtual ward services are not available, the patient would likely remain in hospital, so correct classification of this time is important. Furthermore, anchoring the metric to escalation of care relative to baseline in this way also directly addresses the authors' concern that days-based outcomes do not capture patients' functional status after discharge. Because a patient's location and level of support are themselves coarse but informative markers of their level of dependence, this approach recovers part of this information at a far lower burden, for both patients and investigators, than formal functional assessment. The burden might be further lowered by developing data collection tools, such as an app through which patients or carers record daily location and level of support.

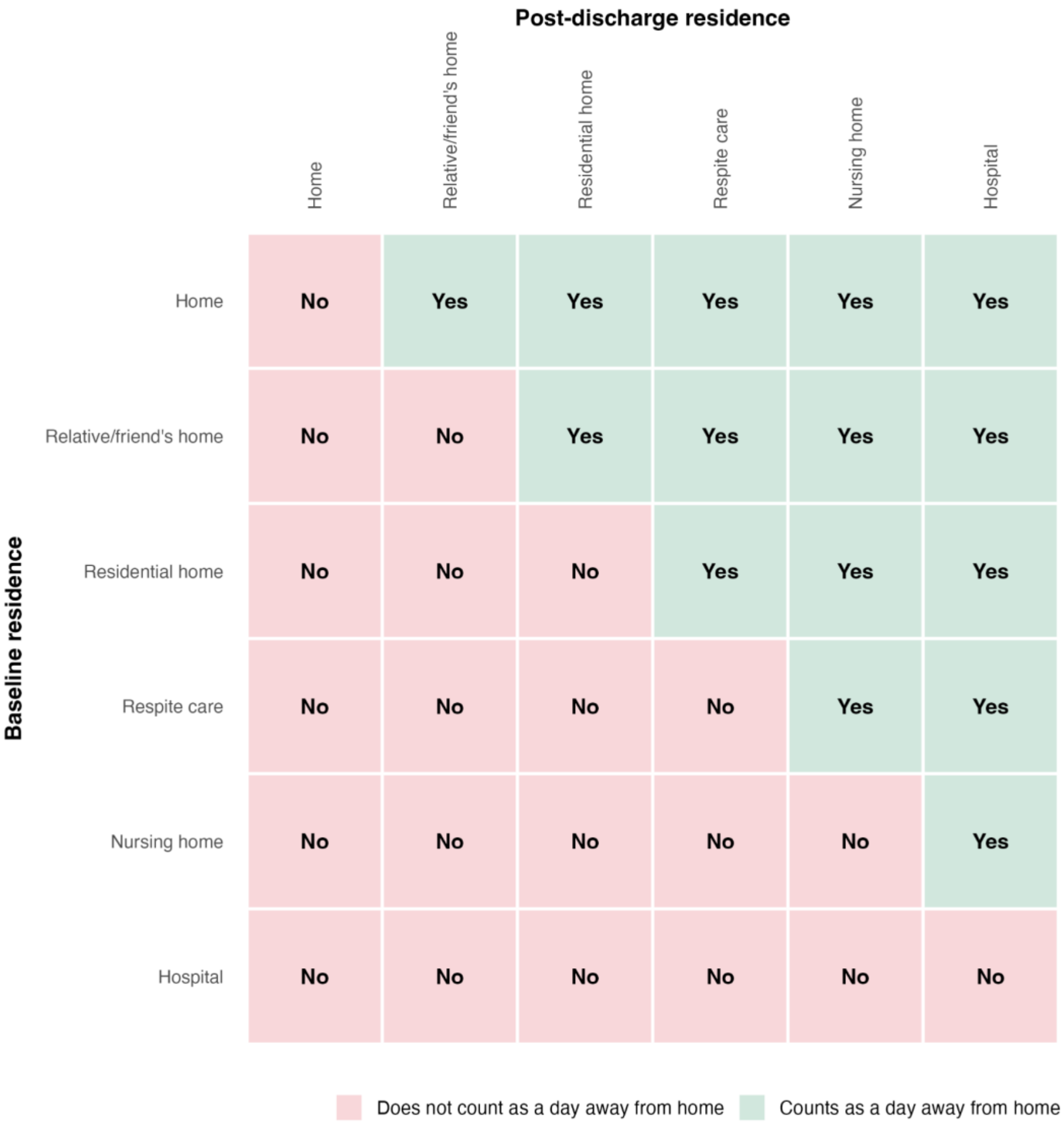

**Figure 2**: What counts as a day away from home in the baseline-adjusted DAH (as used in the NOTACS trial[5])

In summary, the choice of endpoint should follow from the assumed surrogacy between location and recovery, and from where along the patient pathway the treatment effect is anticipated to act. We echo the authors' view that several questions remain open, including the optimal method for incorporating death into these metrics. Our group is currently tackling the methodological complexities of these endpoints—specifically regarding outcome definition and calculation, trial estimand formulation, distribution modelling[9], and imputation methods for missing post-discharge location data[10]. We welcome the authors' call for deeper collaboration and attention to these endpoints, which is vital if we are to successfully standardise their use.

**Funding:** LY is supported by the MRC Trial Methodology Research Partnership Doctoral Training Partnership (TMRP DTP, grant number MR/W006049/1). SND is funded by the Papworth Trials Unit Collaboration. DLC and SSV are partly funded by the UK MRC (grant number MC_UU_00002/1). MST, Advanced Fellow, is funded by the National Institute for Health and Care Research (NIHR) (grant number NIHR305417). The views expressed are those of the authors and not necessarily those of the NIHR or the Department of Health and Social Care.

**Declaration of interest:** The authors have no conflicts of interest to declare, though SSV notes their role on PhaseV's advisory board.